# The Sagittarius impact as an architect of spirality and outer rings in the Milky Way

Chris W. Purcell[1,2], James S. Bullock[2], Erik J. Tollerud[2], Miguel Rocha[2], and Sukanya Chakrabarti[3]

[1] *Department of Physics and Astronomy, The University of Pittsburgh*
[2] *Center for Cosmology, Department of Physics and Astronomy, The University of California, Irvine*
[3] *Department of Physics, Florida Atlantic University*

**Like many galaxies of its size, the Milky Way is a disk with prominent spiral arms rooted in a central bar[1], although our knowledge of its structure and origin is incomplete. Traditional attempts to understand the Galaxy's morphology assume that it has been unperturbed by major external forces. Here we report simulations of the response of the Milky Way to the infall of the Sagittarius[2] dwarf galaxy (Sgr), which results in the formation of spiral arms, influences the central bar and produces a flared outer disk. Two ring-like wrappings emerge towards the Galactic anti-Center in our model that are reminiscent of the low-latitude arcs observed in the same area of the Milky Way. Previous models have focused on Sgr itself[3,4] to reproduce the dwarf's orbital history and place associated constraints on the shape of the Milky Way gravitational potential, treating the Sgr impact event as a trivial influence on the Galactic disk. Our results show that the Milky Way's morphology is not purely secular in origin and that low-mass minor mergers predicted to be common throughout the Universe[5] probably have a similarly important role in shaping galactic structure.**

To discern the specific effect of the Sgr impact on the Galactic disk, we need to simulate directly the dark matter and stellar components in both the Milky Way and the Sgr progenitor and to ensure that Sgr has a realistic dark-to-baryonic mass ratio, given the $\Lambda$CDM (where $\Lambda$ represents the accelerating expansion of the Universe, which has a matter density dominated by Cold Dark Matter) cosmology's prediction that even small dwarf galaxies are hosted by massive halos of dark matter. Given the total luminosity (~ a few times $10^8$ $L_\odot$, where $L_\odot$ is the solar luminosity) of the Sgr tidal stream and remnant core[6], cosmological abundance matching demands that the original mass of the dwarf galaxy progenitor was at least $10^{10.5}$ $M_\odot$ (where $M_\odot$ is the solar mass), although best estimates[7,8] place it in a much more massive halo of roughly $10^{11}$ $M_\odot$. A recent dynamical analysis finds comparable masses, noting that the future discovery of additional stellar debris in the Sgr stream would tend to support the heavier value[6]. We therefore adopt the two masses mentioned above as lower and upper limits, which we refer to as our *Light Sgr* and *Heavy Sgr* models. Our initial Milky Way disk model[9] matches theoretical expectations and observed characteristics of the Galaxy.

In isolation, our modelled primary disk begins to form a weak bar after about two billion years, but otherwise remains quite smooth beyond about 5 kpc from its center. In contrast, the Sgr interactions provide significant perturbations to the outer disk, triggering the formation of outer rings of stellar material and influencing the evolution and formation of the central bar and inner spirality. Each of the model Sgr progenitors experiences two disk crossings, approaching a third at the present day, and the response of the disk is similar in both cases as shown in Fig. 1. The satellite first crosses the disk at a Galactocentric distance of ~ 30 kpc approximately ~1.75 Gyr ago, producing the most significant perturbation. The progenitor loses roughly 75% of its dark matter mass (but little stellar material) during this time, and the disk experiences a caustic signature initially pointing towards the encounter but eventually shearing into trailing spiral ring-like structure; see Fig. 2. The second crossing incites a weaker ancillary arm with a pitch angle different from that of the primary mode and begins to liberate stellar material from Sgr. These repeated polar encounters produce flaring, asymmetric sloshing in the disk plane, and vertical oscillations above and below the plane of the forming spiral wraps.

The evolution of the central bar can also be affected by perturbing impacts. Although bar formation is sensitive to initial conditions, it is interesting to compare results from run to run, which rely on identical primary disk models. Compared to our isolated run, the *Light Sgr* model induces a more pronounced bar with a faster angular speed. Our *Heavy Sgr* case suppresses bar formation compared to the isolated run, as a result of enhanced central disk heating. Though the bar grows with time in the isolated case, at fixed time the *Light Sgr* run always produces a more pronounced bar and the *Heavy Sgr* run always produces a less pronounced bar. Both Sgr-infall models each have an endstate bar orientation ($\phi_{bar} \simeq 15 - 20°$) that corresponds to estimates[10] of the long bar at the center of the Milky Way ($\phi_{MW} \sim 15 - 30°$). Our isolated run does not, being phase-shifted from the impacted bars by roughly 90°, which indicates that the Sgr event must be considered in any model that attempts to detail the evolution of the Galactic bar.

A vital test of the model's viability is the preservation of a disk that is as thin and dynamically cold as the Milky Way. Though our resultant disks do show flaring at large radius, the scale heights remain less than 0.5 kpc well beyond the solar radius in both cases. The velocity ellipsoids of the remnant disks in the solar vicinity are ($\sigma_R$, $\sigma_*$, $\sigma_z$) $\simeq$ (37, 27, 20) km s$^{-1}$ for the *Light Sgr* case and (33, 42, 22) km s$^{-1}$ for the *Heavy Sgr* case, which are grossly consistent with constraints[11] placed on nearby stars with age ~ 4−8 Gyr, i.e. ($\sigma_R$, $\sigma_*$, $\sigma_z$) $\simeq$ (35, 25, 20) km s$^{-1}$.

At the present time in the real Galaxy (and in the infall models, as shown in Fig. 3 for *Light Sgr*), the Sgr core is moving up toward the Galactic plane[12] and has two distinct tidal arms resulting from its advanced stage of disruption. The orbital plane of Sgr allows us to fix the Galactic longitude of the Sgr remnant in our simulations at $l \sim 5.6°$, and we establish the endstate timestep when the dwarf core is at a Galactic latitude of $b \sim$ -14.0°. These coordinates are a good match to observations[2,13]; the heliocentric distance of our satellite remnant is ~22 kpc for the *Light Sgr* model and ~20 kpc for the *Heavy Sgr* case, commensurate with the 24 ± 4 kpc range typically derived[14]. The stellar velocity dispersion of both the core and the stream in our remnants are consistent with measurements for Sgr[5,15] (~ 10 - 20 km s$^{-1}$) though precise results are sensitive to stellar initial conditions. Our simulated Sgr debris distributions do not precisely match all of the observed characteristics, but we argue that these differences are not significant enough to alter our gross expectation that the Sgr impact has significantly affected the Milky Way disk, given that dark matter in the progenitor is the main driver of disk perturbations. Better constraints on debris stream dispersion, length, and thickness may provide a way of constraining the full progenitor mass in the future.

The disks in our simulations develop outer arcs of material generated in association with each disk crossing. These evolved outer wrappings are loosely-wound and resemble rings. One of the predicted arcs at ~10 kpc from the Sun is reminiscent of the low-latitude Milky Way feature known as the Monoceros Ring (MRi). Though MRi is often considered to be the leftover tidal stream from a now-defunct dwarf satellite galaxy[16], some observational evidence has suggested that the MRi could be a feature of the Milky Way itself[17]. Previous theoretical work has suggested that a past encounter with some heretofore-unidentified massive satellite could have produced the MRi as the outcome of a disk impact[18]. We specifically identify the Sgr progenitor as the likely candidate for the impact that molded the Monoceros ring from the Milky Way disk, as the induced spiral arms detached from the outer Galactic plane and began to oscillate vertically over a range of ~5 - 10 kpc (see Fig. 4).

Finally, we note that these predicted ring-like wrappings of already-known spiral arms will also be potentially observable at deep magnitudes by next-generation mapping surveys. These efforts may connect the features in the Galactic outskirts to the global structure of the Milky Way disk, further implicating the Sgr dwarf as a principal shaper of Milky Way morphology. Cosmological considerations strongly suggest that the Sgr progenitor was massive, and thus motivate the expectation that it has influenced Galactic evolution. More broadly, the implication that Sgr has affected the Milky Way morphology provides an indication that minor mergers shape galactic structure throughout the universe. Future observations of the kinematics and extent of the Sgr tidal debris stream will further constrain the scenario discussed here and shape the perspective that the Sgr impact must be included in future theories of Milky Way evolution.

**Supplementary Information** is linked to the online version of the paper at www.nature.com/nature.

**Acknowledgements** We would like to thank Kathryn Johnston, David Law, Heather Morrison, and Andrew Zentner for useful discussions, along with Curt Struck and an anonymous referee for instructive suggestions that improved the final form of this work.  All simulations were performed on the GreenPlanet cluster at UC Irvine.

**Author Contributions**  C.W.P. helped to conceive the project, performed and analyzed all simulations, and wrote the majority of the text.  J.S.B. helped to conceive the project, contributed to the analysis, and co-authored the text.  E.J.T. created the 3D visualizations and provided discussion and direction related to observational correlations.  M.R. provided the code that was utilized to initialize the Sagittarius progenitors and insight on how to thereby achieve the desired properties of these systems.  S.C. assisted with interpretation and analysis of the stellar disk instabilities and their time evolution.

**Author Information** Reprints and permissions information is available at www.nature.com/reprints.  Correspondence and requests for materials should be addressed to C.W.P. (cpurcell@pitt.edu).

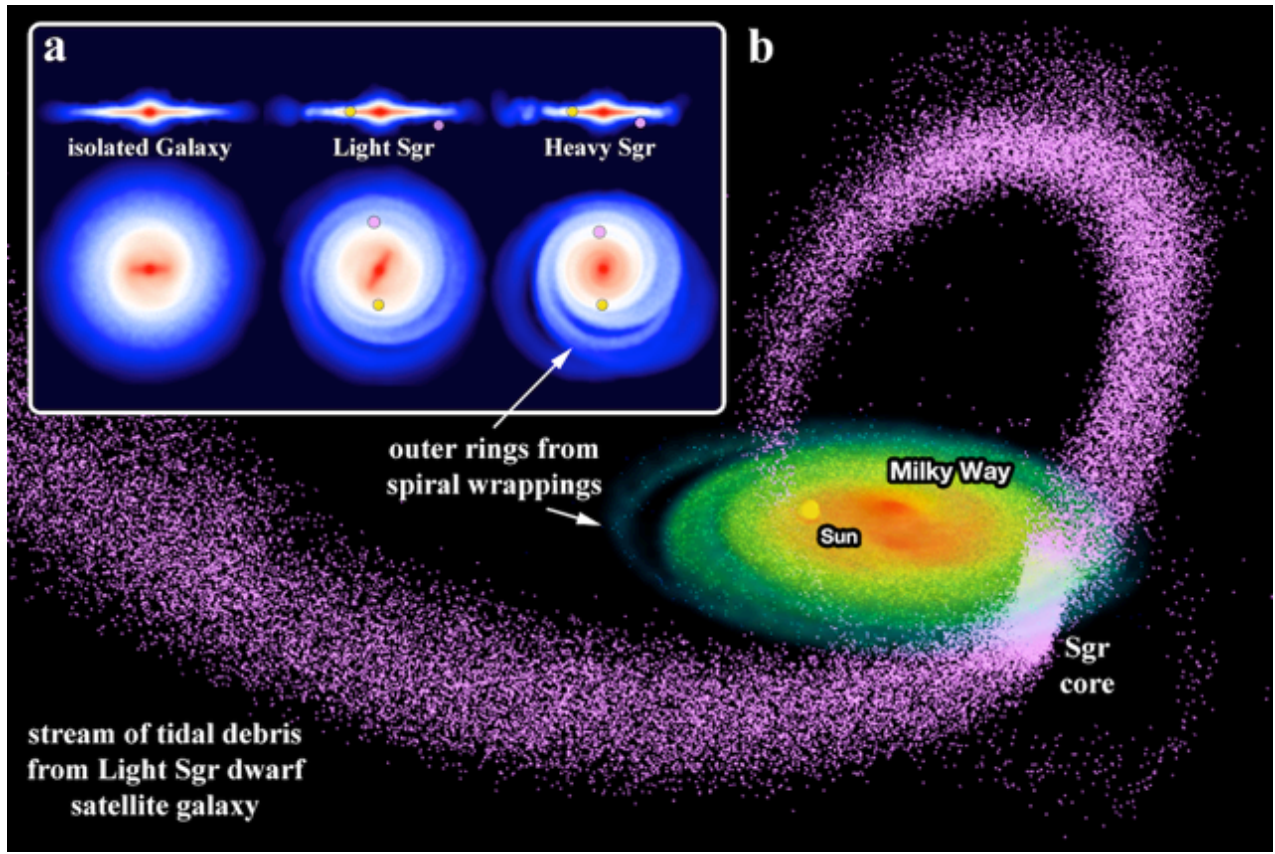

**Figure 1** - **Visualizations of evolved disk endstates in the simulation suite**. **a**, Edge- and face-on surface density depictions for each infall model as well as an isolated Galaxy model subject only to secular evolution. The solar location is marked in yellow and the present location of the Sgr remnant is marked in pink. The primary Milky Way analog was initialized via self-consistent multi-component distribution functions[9] and proved fairly robust to secular instabilities, as shown in the left image after ~2.7 Gyr of isolated evolution. **b,** Global rendering of the *Light Sgr* endstate disk and tidal debris. The primary galaxy included a Navarro-Frenk-White (NFW) dark halo[19] with scale radius $r_s$ = 14.4 kpc and virial mass $M_{vir}$ = $10^{12}$ $M_\odot$; the disk had a mass of 3.59 × $10^{10}$ $M_\odot$, an exponential scale length of 2.84 kpc and a vertical $sech^2$ scale height of 0.43 kpc; the central bulge had a mass of 9.52 × $10^9$ $M_\odot$ and an $n$ = 1.28 Sérsic profile with a 0.56 kpc effective radius. The *Light Sgr* (*Heavy Sgr*) progenitor with effective virial mass $M_{vir}$ = $10^{10.5}$ $M_\odot$ ($10^{11}$ $M_\odot$) was initialized with an NFW dark halo of scale length 4.9 kpc (6.5 kpc) self-consistently with a separate stellar component[20] motivated by an analysis of the observed Sgr debris and core[6]: a King profile[21] with core radius 1.5 kpc, tidal radius 4.0 kpc, and central velocity dispersion equal to 23 (30) km s$^{-1}$. Following previous work on the Sgr interaction[22], our satellites started 80 kpc from the Galactic Center in the plane of the Milky Way, traveling vertically at 80 km s$^{-1}$ toward the North Galactic Pole. We account for the mass loss that would have occurred between virial-radius infall and this "initial" location by truncating the Sgr progenitor NFW mass profile at the instantaneous Jacobi tidal radius, $r_t$ = 23.2 kpc (30.6 kpc), leaving a total bound mass that is factor of ~3 smaller than their effective virial mass from abundance matching. All simulations used the parallel N-body tree code ChaNGa with a gravitational softening length of one parsec, and followed the evolution of 30 million particles with masses in the range 1.1 - 1.9 x $10^4$ $M_\odot$.

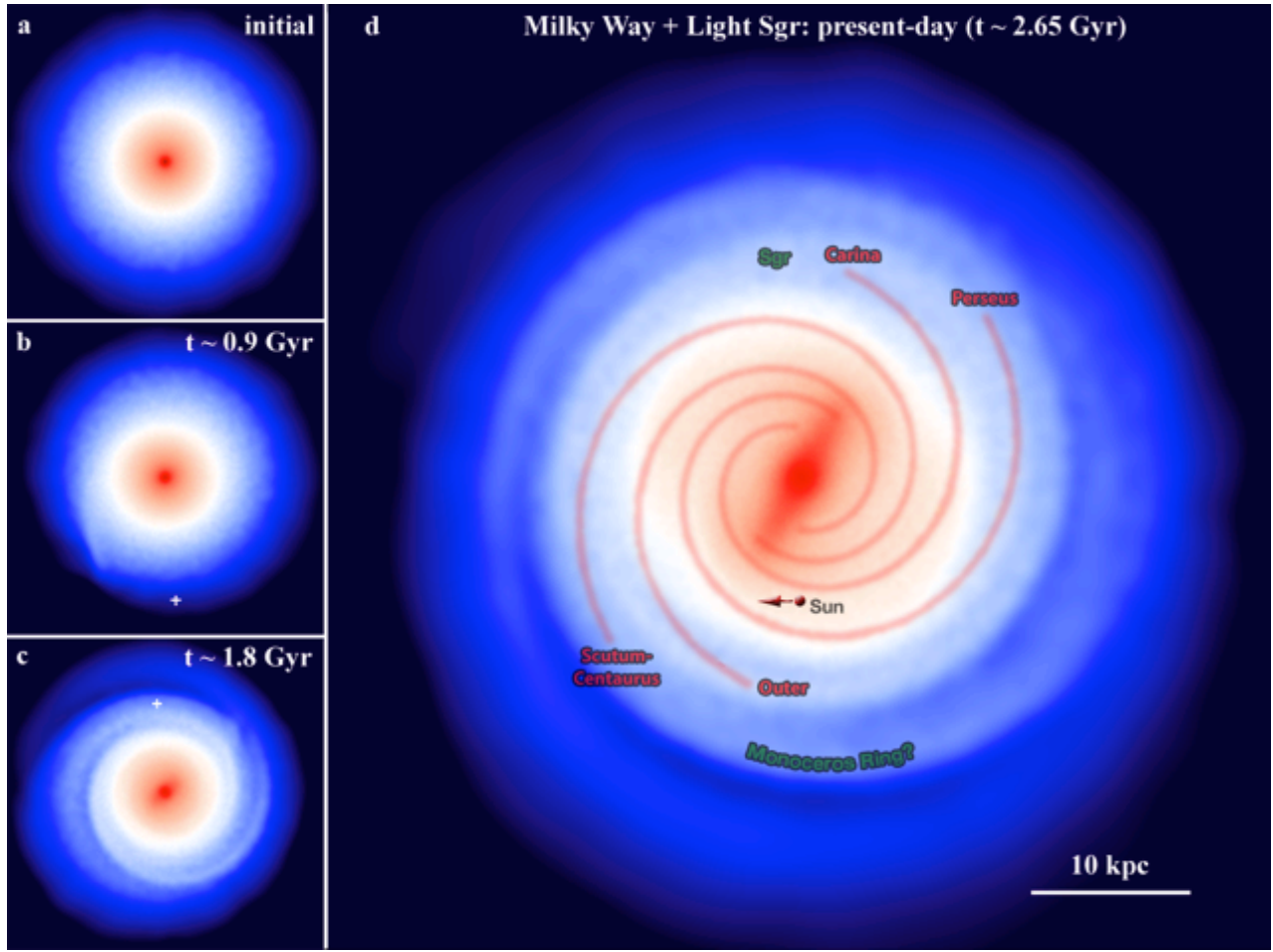

**Figure 2** - **Face-on surface density visualizations of the Milky Way at four important moments during the *Light Sgr* simulation**. **a**, Initial model. **b**, Immediately following first pericenter, where the white cross marks the Sgr impact point. **c**, Shortly after second pericentric disk crossing. **d**, At the present-day (corresponding to elapsed simulation time 2.65 Gyr), overlaid by a four-armed symmetric-spiral fit to the observed arms of the Milky Way as revealed by mapping neutral hydrogen[23]. The traditional view of the Milky Way as a secularly-evolving system has encouraged theoretical descriptions of quasi-stationary density-wave spirality, although the large peculiar motions of young stars in spiral arms support a more transient picture[24] (numerical evidence exists for both short-lived configurations[25] as well as more stable forms of spirality, varying with the strength of the tidal induction[26]). Dynamical analysis of each impacted Milky Way model reveals the importance of the swing amplification mechanism, in which gravitational disturbances in the stellar disk at each pericentric approach shear into trailing arms that are subsequently enhanced on small scales (even in a globally stable system), strengthening transient spiral modes.

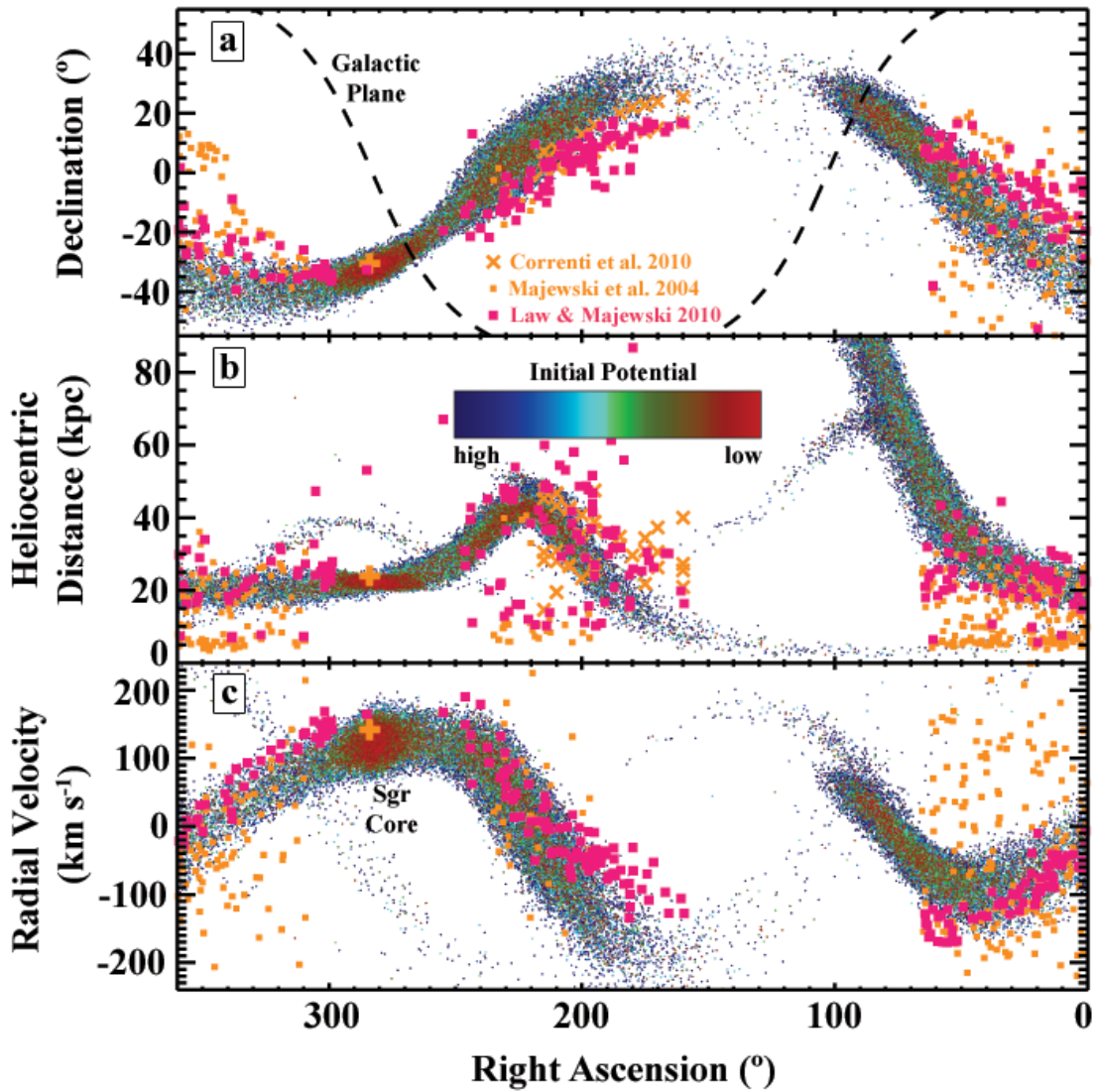

**Figure 3** – **The observed Sgr tidal debris stream and remnant core in comparison to our *Light Sgr* simulation, in equatorial coordinates**. **a**, Declination versus right ascension. **b**, Heliocentric distance versus right ascension. **c**, Radial velocity versus right ascension. Simulated particles are colored according to their initial potential energy, and the orange points are data from 2MASS M-giant stars[27] and SDSS red-clump stars[28] (marked by squares and crosses respectively; thick crosses denote canonical values for the remnant core[2,12]). The pink points are 2MASS M-giants identified as likely stream members[5]. The present-day location of the simulated remnant and tidal arms are similar to those observed Combining this with observational constraints on the dispersion ($\sigma \sim 10 - 15$ km s$^{-1}$), breadth (8 – 10 kpc)[27], and length of the observed debris stream provides some legitimacy for our model.

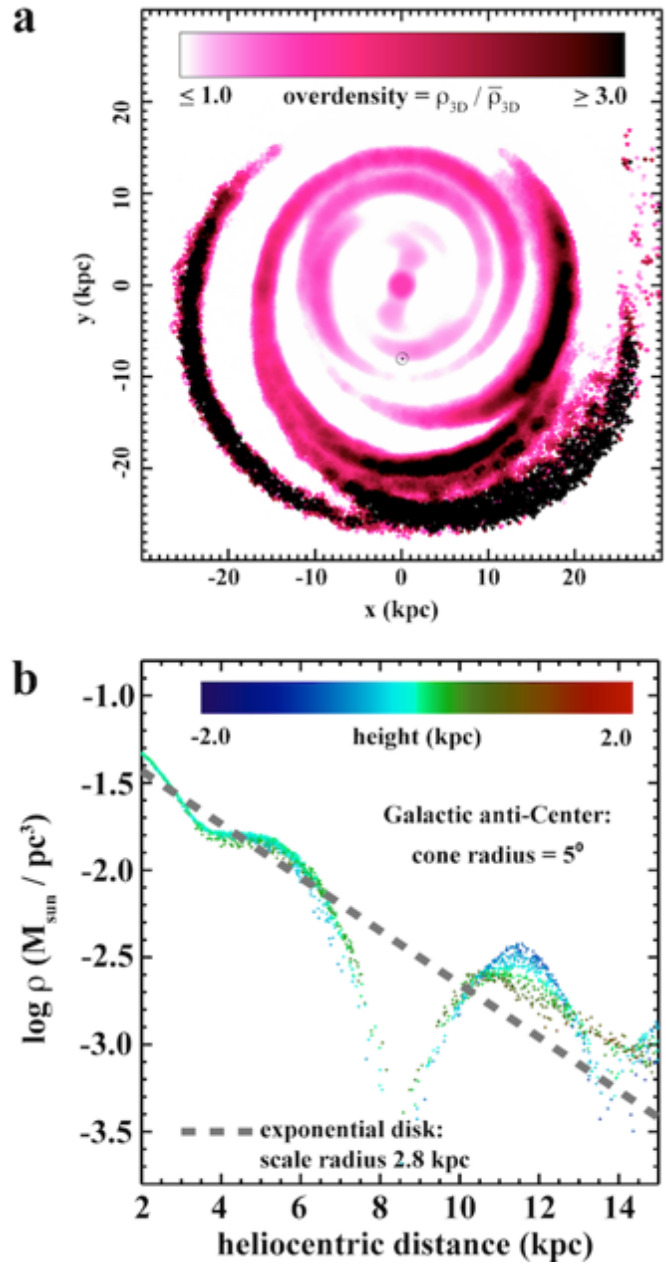

**Figure 4** – Endstate disk overdensities in the *Heavy Sgr* simulation.  **a**, Disk stellar overdensity color-coded by the ratio of 3D stellar density to the local axisymmetric mean density of the disk.  **b**, Local stellar density in a thin cone directed from the solar neighborhood toward the anti-Center, as a function of heliocentric distance along the cone. In both panels, off-plane overdensities resemble "multiple tributaries" observed in the Milky Way[29], and the spiral arm wrapping at a distance of ~10 kpc is strikingly similar to the Monoceros Ring (MRi). The MRi feature spans a wide range in metallicity ([Fe/H] ~ -1.6 to -0.4)[30]. The corresponding high-latitude arc in our simulation at that distance is composed of stars that were initially distributed widely throughout the disk, suggesting that radial mixing during the Sgr impact must be a significant factor in the Milky Way's recent chemodynamical evolution, and also that the chemical composition of the real MRi feature may not be as reliable in discriminating models for the origin of the ring as otherwise expected. See the *Supplementary Information*, where we discuss the quantitative agreement between our anti-Center spiral structure and Monoceros Ring observations in more detail.

## Supplementary Information:

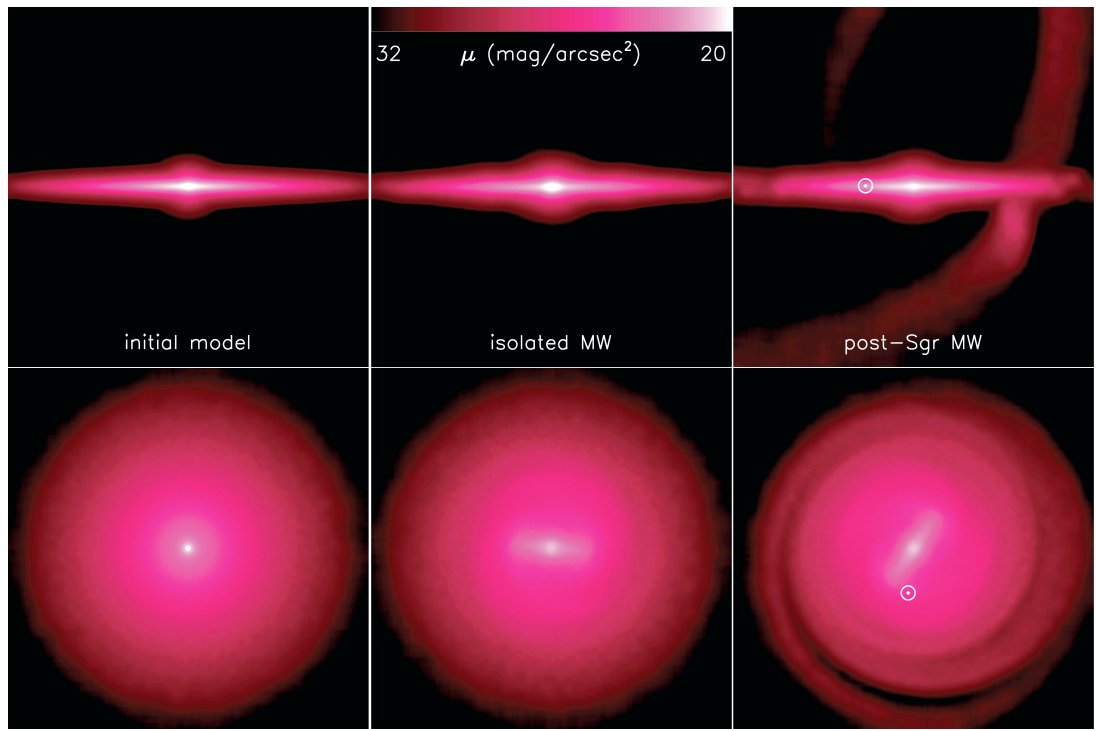

**Figure S1** – Surface brightness maps (assuming a stellar mass-to-light ratio M*/L = 3) for the initial state of our primary Milky Way model (*left* panels) and the endstate following the infall of the *Light Sgr* dwarf galaxy (*right* panels). In the *center* panels, we show the result of secular evolution in the initial model for an equivalent timespan of 2.65 Gyr; note that the Sgr interaction causes the widespread emergence of discernible spiral arms and influences the evolution of the central bar. In the *right* panels, the orientation of the satellite's orbital plane allows us to define the position of the Sun (marked by the solar symbol).

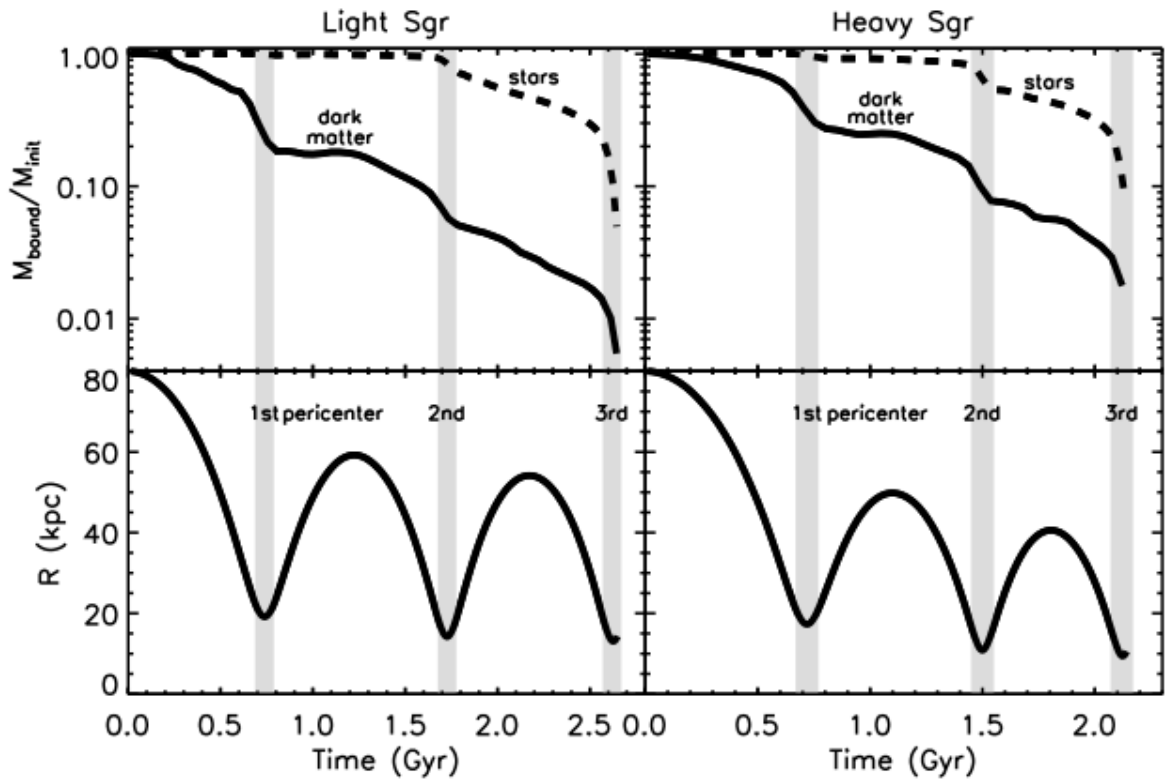

**Figure S2** - The time evolution of bound mass and orbital radius in each simulated Sgr infall. Most mass loss occurs near the pericentric passages, marked by shaded bands. Note that the more massive *Heavy Sgr* model (*right* panel) decays more quickly than the *Light Sgr* model (*left* panel) due to increased dynamical friction and tidal disruption efficiency. The dark halos in these satellite models both follow NFW density profiles with a scale length of 4.9 (6.5) kpc for *Light* (*Heavy*) *Sgr*. We account for the mass loss that would have occurred between virial-radius infall and our "initial" Galactocentric radius of 80 kpc by truncating the Sgr progenitor mass at the instantaneous Jacobi tidal radius at that position, *i.e.* $r_t$ = 23.2 kpc for the *Light Sgr* model and 30.6 kpc for the *Heavy Sgr* case. The total mass enclosed within this radius is $M_{Sgr}$ = 1.37 × $10^{10}$ $M_\odot$ and 3.16 × $10^{10}$ $M_\odot$ for the *Light* and *Heavy* models, respectively. These masses agree well with the pre-disruption mass estimates based on the stellar kinematics of the observed Sgr core and debris stream[6]. Implicit in our model is that the infall into the halo was recent enough that the first pericenter crossing modeled here was indeed the first close passage experienced by this satellite.

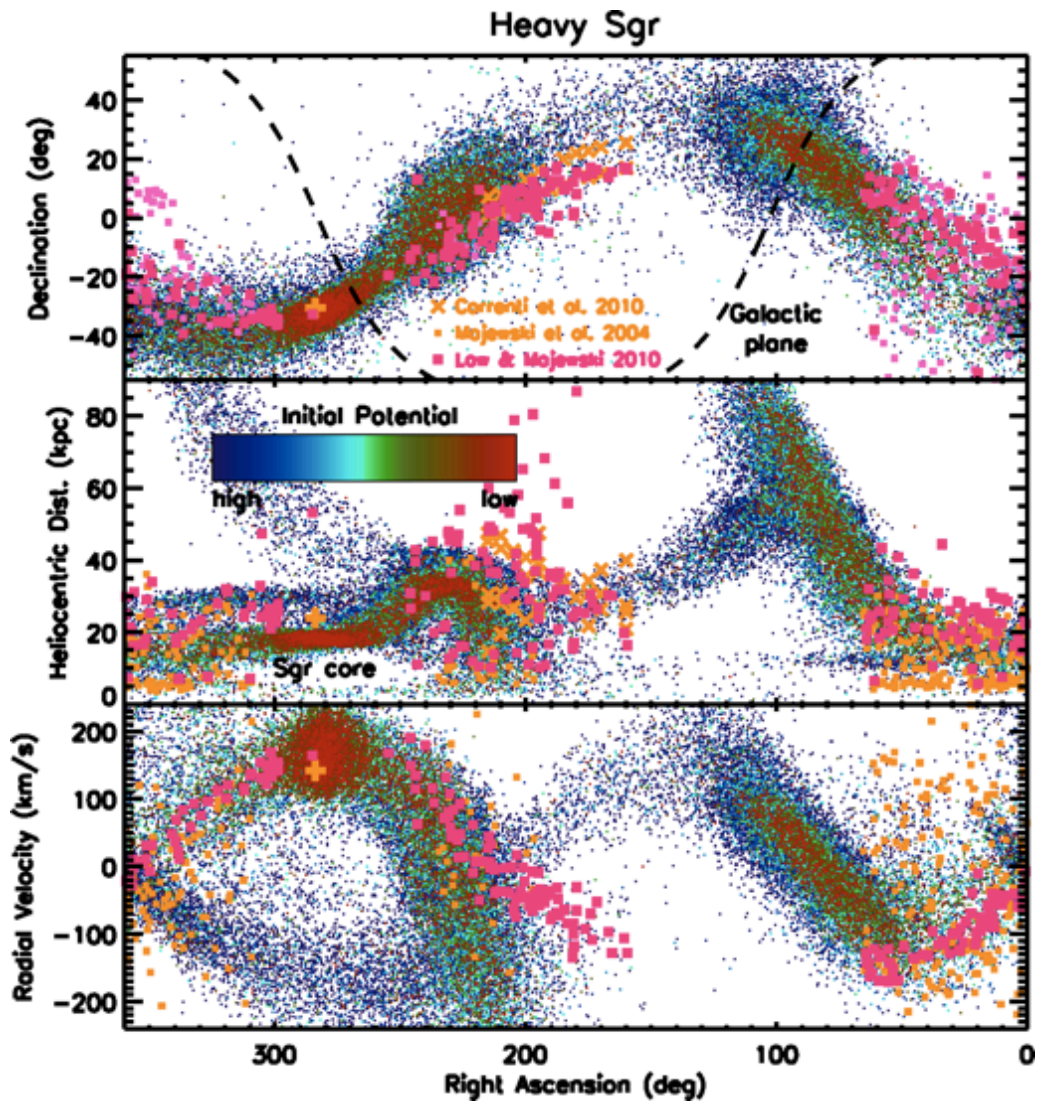

**Figure S3** – Similar to Figure 3 in the main text, the final distribution of stellar material initially belonging to the *Heavy Sgr* satellite, in equatorial coordinates and compared to SDSS[27] and 2MASS[5,28] observations of the dwarf's tidal streams (crosses and squares, respectively). Thick crosses denote the canonical observed values for the Sgr remnant core[2,13].

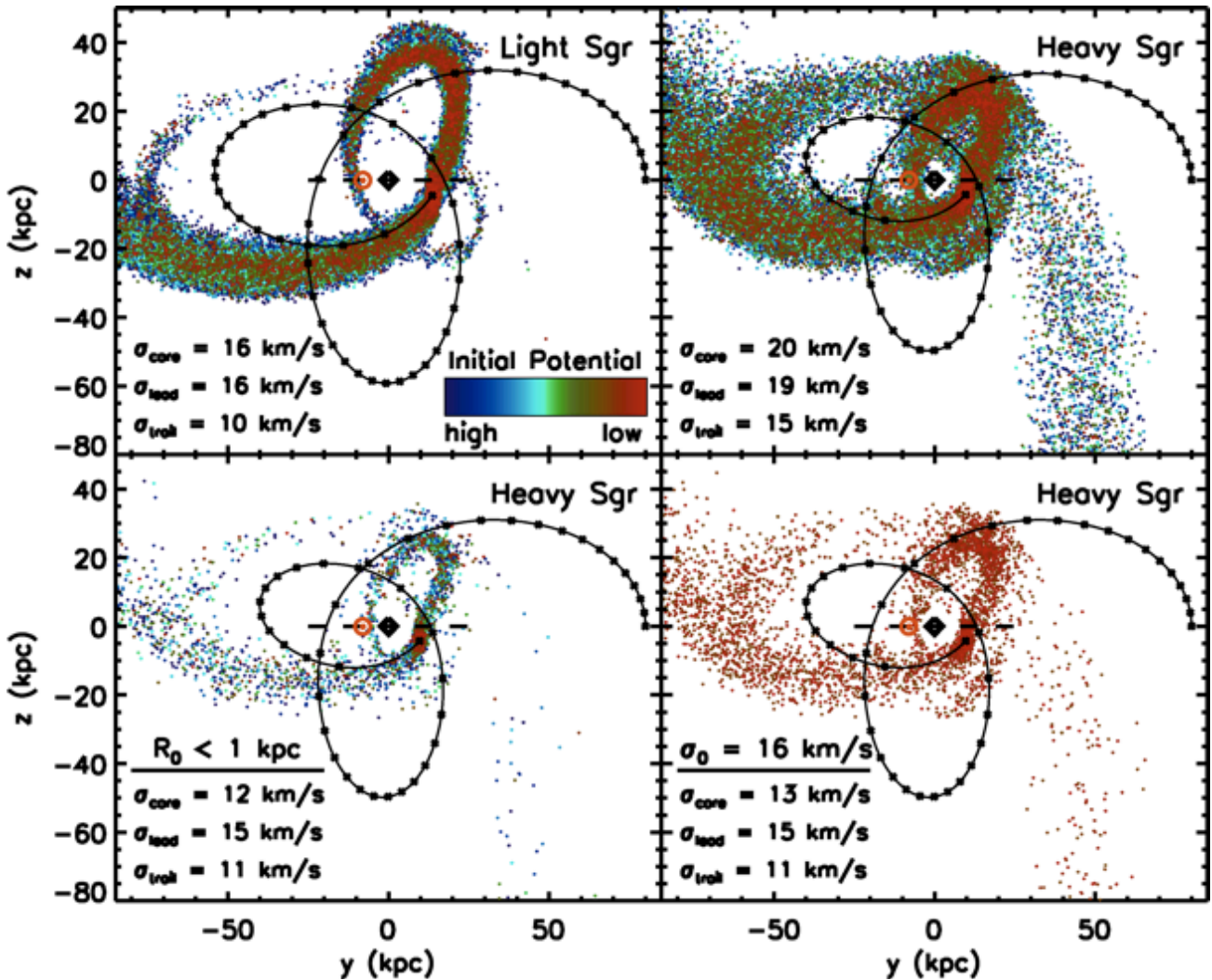

**Figure S4** – Cartesian projections into the orbital plane of the Sgr stream debris for both initial satellite models (*top* panels), and for the *Heavy Sgr* model with stellar subsampling imposed such that the resultant stream kinematics are significantly colder than in the fiducial *Heavy* case (*bottom* panels). The *lower left* panel includes only star particles that were within 1 kpc of the core initially. The *lower right* panel imposed an energy cut on the initial star particles, such that the initial central velocity dispersion matched that of the fiducial *Light Sgr* model (16 km/s). Resultant velocity dispersions for the core, leading, and trailing arms are listed in each panel. There are clearly significant degeneracies between the total dark matter mass and the initial stellar distribution within the progenitor. In order to use the length and kinematics of the debris stream as a means to constrain the progenitor mass, one would need to explore a multi-dimensional parameter space that includes the initial conditions of stars within Sgr. Future work in this direction is warranted.

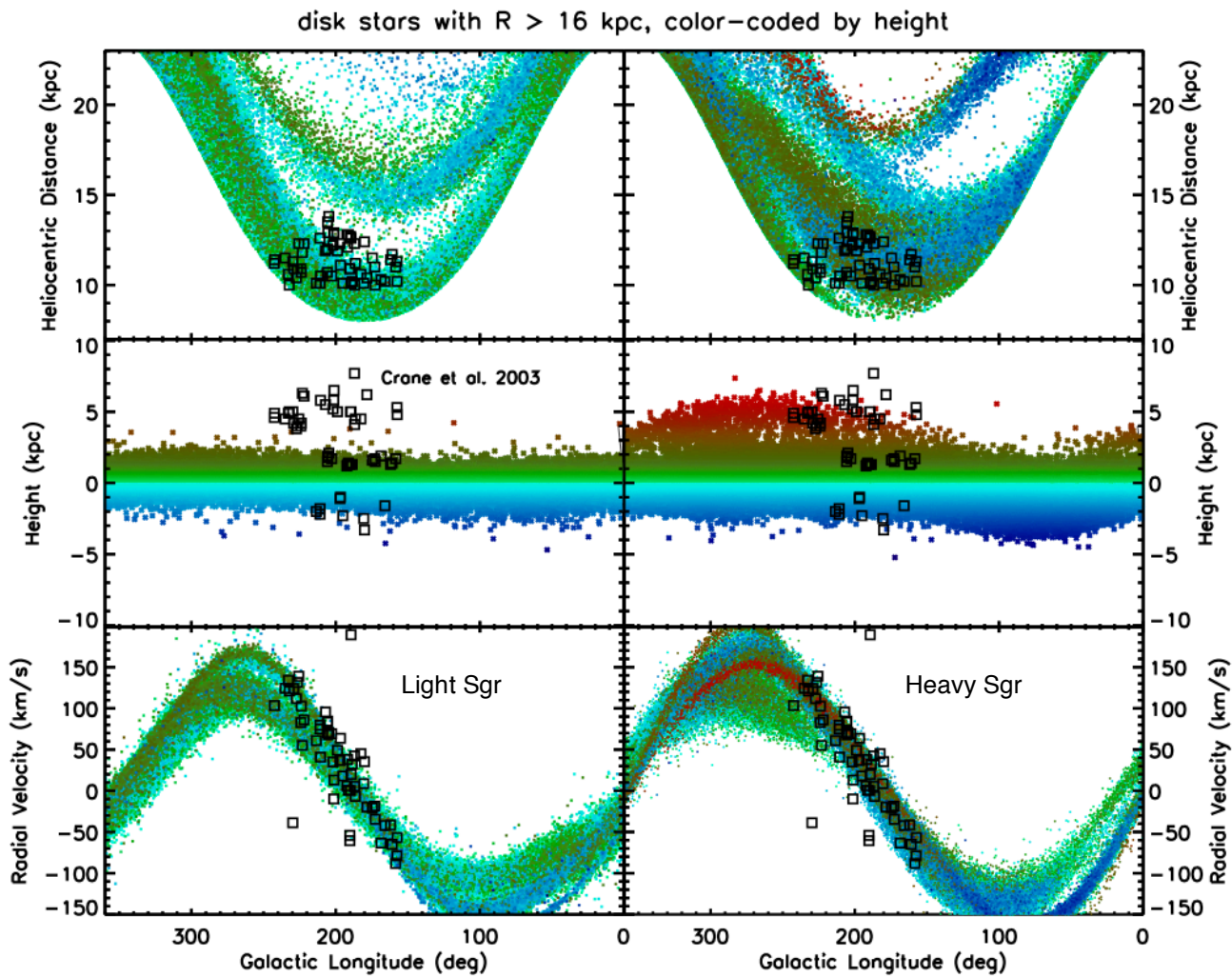

**Figure S5** - The final properties of disk stars with radius R > 16 kpc for the *Light* and *Heavy Sgr* models (*left* and *right* columns, respectively), with data points drawn from the Galactic anti-center region[31]. Note that in all panels, the disk star particles are colored according to their height above or below the Galactic plane.

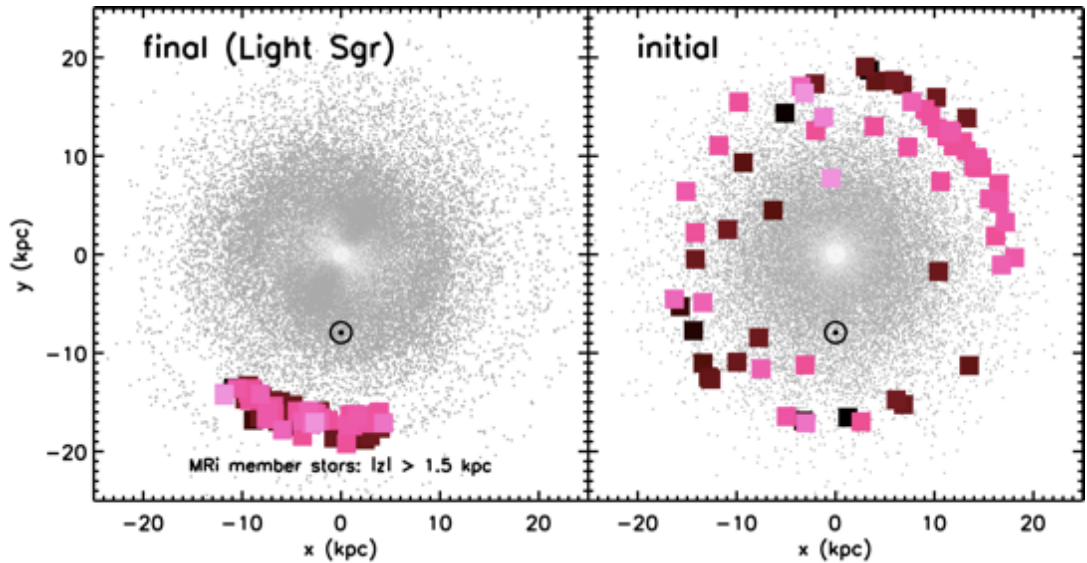

**Figure S6** - The distribution of high-latitude star particles in the MRi-like feature, for the endstate (left) and initial disk (right) in the *Light Sgr* simulation. These MRi member stars are colored by their final distance from the mid-plane, where the darkest (lightest) particles correspond to ~1.5 kpc (~3.0 kpc). For reference, a selection of disk particles are plotted in the background, and the solar position is marked in each panel. Note the high degree of radial mixing that has occurred, with the MRi feature at present day being composed of disk stars from a wide range of initial positions.

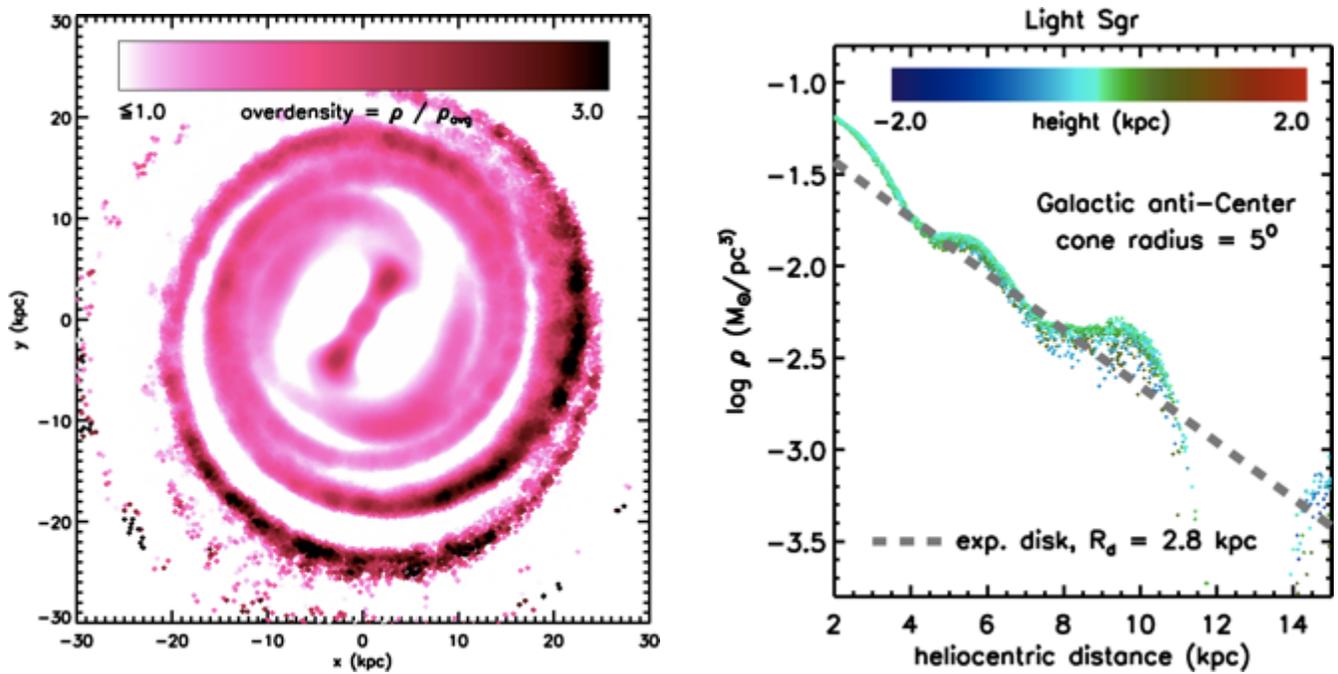

**Figure S7** – Similar to Figure 4 in the main text, visualization of the global disk structure in our *Light Sgr* simulation. The *left* panel presents an over-density map of the present-day stellar disk and the *right* panel shows the local density measured in a 5-degree cone emerging from the solar position and directed toward the Galactic anti-Center.

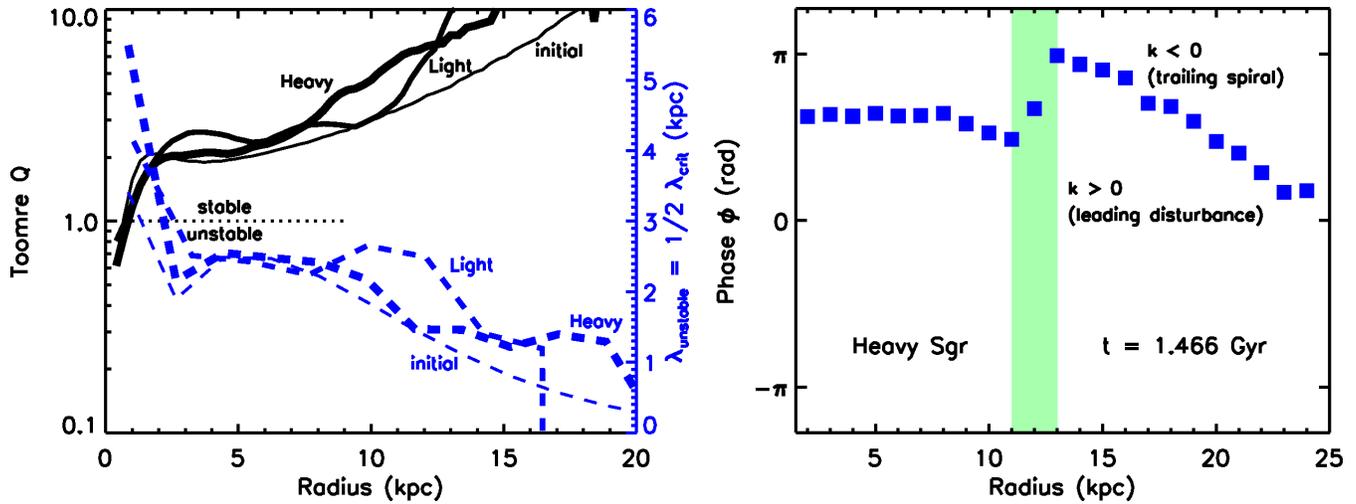

**Figure S8** - *Left* panel: the Toomre stability parameter Q and the critical wavelength associated with scales most susceptible to instability. Where necessary, we set the mode number *m*=2 for two-armed spiral and bar structure. Note that the initial Milky Way model, as well as both endstate disks from the *Light Sgr* and *Heavy Sgr* infall models, are globally stable with Q > 1 at all radii where disk particles dominate the stellar density. Small-scale perturbations with wavelengths less than roughly 2-3 kpc travel easily through all of these disks. *Right* panel: the phase (or shape function) $\phi$ as a function of radius, during the second pericentric approach/impact of the *Heavy Sgr* satellite, and annotated regionally with values of radial wavenumber $k = d\phi / dR$. Near the radius of disk impact (shaded area), a leading density disturbance emerges due to the gravitational perturbation of the interloping subhalo, and this feature is quickly sheared by differential rotation into the generally-trailing spiral structure of the post-infall disk as time evolves.

We visualize the stars associated with the primary Milky Way model in the *left* panels of **Figure S1** and the same system evolved in isolation for 2.65 Gyr (the time of the longest Sgr orbit simulation) in the *middle* panels. We note that these initial conditions are quite stable during secular evolution, though a weak bar does begin to emerge after ~2 Gyr of evolution. The corresponding evolutions of the *Light* and *Heavy Sgr* progenitor models are presented in the *left* and *right* panels of **Figure S2**, respectively. The upper panel shows the bound mass in each component in units of the initial mass, while the lower panel shows the radial evolution of each orbit. Each of the modeled Sgr progenitors experiences two disk crossings and approaches a third, corresponding to the three orbital pericenters (as indicated by shaded bands). For both models, the first impact occurs in the outer disk at $R_{GC}$ ~ 30 kpc and is responsible for most of the dark matter mass loss in the satellite but very little stellar mass loss. The second occurs ~0.9 Gyr later at $R_{GC}$ ~ 15 kpc and is responsible for liberating the majority of the stellar stream debris. The final disk crossing is occurring at the present day, about ~1.8 Gyr after the first disk crossing. We note that the impact times and radii for these models are quite similar to those presented in past models aimed at reproducing the Sgr stream in great detail[5].

The observed features of the Sgr stream[5,27,28] are reproduced well in the endstate of our experiment, as indicated in **Figure S3** (for the *Heavy Sgr* model; see Figure 3 in the main text for the corresponding *Light Sgr* result). Though our simulated Sgr debris distribution does not precisely match all of the observed characteristics, we claim that these differences are not significant enough to alter our gross expectation that the Sgr impact has significantly affected the morphological evolution of the Milky Way disk. These differences in the stellar debris are likely unimportant compared to dark matter in the progenitor, which is the main driver of disk perturbations. Along these lines, we note that the above observational correlations regarding the satellite's tidal stream and remnant core also hold grossly true in the *Heavy Sgr* simulation, but with a debris population that is slightly too hot compared to current estimates. Note that this does not invalidate the claim that the Sgr progenitor was initially very massive, it simply indicates that the initial stellar distribution we chose was somewhat too hot in the case of *Heavy Sgr*. The dynamical response of the disk is dominated by the total mass of the progenitor, not the details of the stellar distribution, which only accounts for ~1-5% of the system mass at first impact. In other words, the current stream could have originated from a wide variety of initial stellar distributions without significantly affecting the overall impact of the progenitor dark subhalo on the Milky Way disk.

We demonstrate the degeneracy associated with stellar distribution initialization in **Figure S4**. In the top two panels we show Cartesian projections of the stellar stream material in our standard *Light Sgr* and *Heavy Sgr* models. In the bottom two panels we show only a subsample of the *Heavy Sgr* star particles, thus effectively treating some fraction of the initial star particles as dark matter. In the lower left panel we redefine the stellar material to the subset that was initially within 1 kpc of the progenitor core. In the lower right panel we have instead imposed a cut on energy, effectively lowering the central velocity dispersion to that of our standard *Light Sgr* case. Resultant velocity dispersions for the core, leading, and trailing streams are indicated. Note that in both resampled

*Heavy Sgr* cases the resultant streams are colder than the original (by design), with similar velocity dispersions to that of the *Light Sgr* model. However, the energy-cut model is thicker, resembling the spatial structure of the original *Heavy Sgr* case. The spatially trimmed model is more reminiscent of the original *Light Sgr* experiment. Overall, this exercise demonstrates that it will be useful to explore the degree to which the length, thickness, and kinematic temperature of the real Sgr debris stream can be used to place constraints on the initial dark matter mass of the Sgr progenitor. However, a multi-dimensional parameter exploration will be essential for discerning the degeneracy between overall dark matter mass and stellar distribution.

A series of additional uncertainties and limitations in our initial conditions make a detailed match of Sgr stream observations somewhat beyond the scope of this current work. Though we do manage to get the gross properties of the stream and core correct, we cannot expect to achieve perfect agreement under the limitation of a spherical Milky Way dark matter halo. As demonstrated by many previous authors[5,33,34], a non-spherical dark matter distribution is likely essential for reproducing the full spatial and velocity structure of the stream. Moreover, from our preliminary experiments, only minor tweaks to the initial location and velocity of our Sgr progenitor resulted in significant differences in the final position and velocity of the debris and core. These details, while important, are beyond the scope of this work, which aims primarily to make the point that the Sgr impact likely influenced Milky Way disk evolution in significant ways.

In our simulations of the Sgr accretion, outer arcs of the MRi type are generated naturally in association with each Sgr disk crossing. **Figure S5** shows a comparison of our resultant outer ring wrappings in each simulation to a typical kinematic MRi sample[31]. In both cases, the rings are fairly cold kinematically and consist of material that resides both above and below the disk. The *Heavy Sgr* model is more capable of generating the amplitude of vertical oscillation necessary to push ring material to the heights observed in MRi candidate stars. However, in much the same way that Sgr stream properties are sensitive to progenitor initial conditions, the vertical extent of heated disk stars will also be sensitive to the uncertain initial vertical disk structure of the primary. For example, had we used an exponential vertical disk structure as opposed to a $\mathrm{sech}^2$ distribution then the vertical restoring force would have been less powerful (for the large-z behavior), thus producing a more vertically extended remnant. We return to this point briefly below.

There has been some concern in the literature that the MRi cannot be associated with disk material because of its enhanced metallicity[16]. **Figure S6** illustrates that this is not a serious concern in the case of our simulations. Disk particles that we identify as high-latitude MRi-analogue stars (magenta points, *left* panel) come from a wide range of initial radii within the disk (*right* panel). This implies that a wide range of metallicity and α-abundances in anti-Center wrappings is likely. Future low-latitude surveys of the outer disk should provide a useful test of this scenario, although great care must be taken to chemodynamically distinguish disk populations from possibly-accreted interlopers, given the relatively low stellar densities in the excited rings as shown in the *right* panels of **Figure S7**.

The *left* column of **Figure S7** visualizes the global disk response to the Sgr interaction in the *Light Sgr* by using a face-on disk rendering color-coded by ratio of the local density to the radial/vertical average at each point. Under-densities are white in this depiction, while over-densities are colored. The images reveal a prominent bar as well as a complicated array of spirality that reveals itself in the nearby anti-Center as overdensities rotating coherently with the Milky Way disk. As mentioned above, the *right* column depicts the local density as determined within a 5-degree solid angle cone along the plane as a function of radius from the solar position, in the direction of the anti-Center region.

In order to quantify the presence of spiral arms in the resultant disk of the fiducial infall model, we have examined the Fourier amplitudes of the final density distribution, finding significant power in the *m*=2 and *m*=4 modes (corresponding to two-armed and four-armed morphologies respectively), with comparatively negligible power in the first two odd modes relative to axisymmetry. This is broadly consistent with analysis of satellite impacts onto the Milky Way[32], which finds that satellites with mass ratio ~ 1:50 and orbital pericenter distances ~ 20 kpc do induce power in the first few Fourier modes, with comparatively greater amplitude in the *m*=2 mode. Analysis of the Toomre stability Q parameter (which similarly assesses axisymmetric disk instability; we encourage the reader to consult the standard dynamical literature[35] for details regarding this type of analysis) and associated critical wavelength $\lambda_{crit}$ indicates that our fiducial Milky Way model is globally stable against long-wavelength perturbations, but is susceptible to short-wavelength modes on small scales, as demonstrated by **Figure S8**. We defer a full stability analysis of the non-axisymmetric disturbances (as has been performed for self-gravitating disks[36]) to a future paper.

We also postpone for future investigation the role of Sgr in producing spirality in the outskirts of the HI disk of the Milky Way, since such a program would require a full hydrodynamical treatment of the Sgr-MW interaction with comparable resolution to that considered in the present work. One consistent finding from a wide range of hydrodynamical studies is that disturbances in the gaseous disk dissipate on the order of a dynamical time[37]. Given that the last Sgr disk crossing occurred roughly a Gyr ago, a large amount of power may not remain in the Fourier modes of the HI disk at the present day. Nonetheless, this is speculative and needs to be confirmed by simulation. It is likely that, at the very least, the Sgr-MW interaction will introduce some power in the Fourier modes even at the present epoch (particularly in the type of paradigm more well-described by the *Heavy Sgr* model), and this needs to be understood in order to more properly interpret the observations of the Galactic HI disk. Our work here compels a more detailed treatment of the Sgr-MW interaction's ramifications for the HI disk, and we leave quantification of the various possibly-competing effects to a forthcoming paper. In addition, the presence of other known (in the case of the Magellanic Cloud system, *modulo* orbital concerns[38]) and hitherto-undiscovered (as in the case of the inferred massive dark subhalo that may have warped the neutral gas in the outer disk[32]) satellites must also be included in future work that seeks to understand the nuances of the Milky Way's recent dynamical response, especially as measurements of the Magellanic Clouds' proper motions begin to constrain that subhalo's history with respect to the overall Galactic evolution.

Finally, we suggest that the observable effects of the Sgr event may well be more obvious in the real Milky Way than they were in our simulation. For example, flaring in the outer disk is highly dependent on the vertical self-gravity of the system, and this gravitational resistance is determined largely by the stellar density distribution very near the disk's mid-plane[39]. Our initial conditions use a $sech^2(z)$ density profile, which compiles relatively more stellar material into the mid-plane region compared to an exponential disk, which seems to be a better match to the Milky Way in detail[40]. We thus posit that the flaring of our fiducial disk model during satellite infall is a firmly conservative estimate, which may well explain the discrepancy shown in the middle panels of **Figure S5** between our result and that subset of MRi member stars observed at particularly high latitudes, given that a decrease in self- gravitational resistance (or equivalently, an increase in the perturbative force as in our test case) can apparently eject disk stars to comparable heights as the outer spiral-arm wrappings detach from the Galactic plane.

One corollary to this behavior draws special significance from the state of current observation regarding the tidal streams extending from the Sgr dwarf. The progenitor re-assembly discussed above[6] comes with an important caveat: when drawing the connection between present-day luminosity and progenitor mass, the discovery and consideration of stellar material as yet hidden by the tidal streams' bunching at multiple orbital apocenters would quickly increase the virial mass predicted for the infalling Sgr dwarf. Only one debris wrapping has thus far been investigated in this way, and it therefore appears quite likely that the Sgr progenitor mass we consider in this project is in fact quite a conservative estimate, to be refined as next-generation surveys provide unprecedented constraints on the Sgr stream's distribution at large Galactocentric radii.

The imminent advances in mapping the Milky Way environment will stimulate much- needed future efforts to refine and expand simulation as it approximates the real universe ever more precisely. The collisionless nature of our simulated disk-satellite impact may well underestimate the efficiency of spiral arm formation via swing amplification; this specific behavior is a transient phenomenon since the increase in random stellar motions self-limits the growth of spirality and dampens the responsiveness of the disk, if there is no fresh infalling gas to replenish circularity by dissipating random motions via collision as star formation proceeds[25]. In addition, the complex dynamical coupling between gaseous and stellar components creates a feedback cycle in which preferred resonance modes are selectively reinforced. The critical wavelength that determines the characteristic scale of the stellar response is significantly larger than the typical scale of gas perturbations, and thus the stars augment the amplitude and the scale of this noise induced by gravitational disturbance as the Sgr subhalo impacts the disk[41] (according to detailed treatments of swing-amplification as it relates to the excitation of intermediate-scale spiral structure[35]). It appears quite likely, all things considered, that the addition of fully-realized hydrodynamical physics to our simulation could enhance the spirality engendered by the Sgr infall.